\documentclass{llncs-zhaoyw-modified}

\usepackage{llncsdoc}
\usepackage{mathptmx}
\usepackage{latexsym}
\usepackage{amsfonts}
\usepackage{amssymb}
\usepackage{graphicx}
\usepackage{color}
\usepackage{multirow}
\usepackage{wrapfig,lipsum,booktabs}
\usepackage{semantic}
\usepackage{isabelle,isabellesym}

\newcommand{\sectprefix}{{Section}}
\newcommand{\subsectprefix}{{Subsection}}
\newcommand{\figprefix}{{Fig.}}

\newcommand{\tableprefix}{{Table}}

\newcommand{\lemmaprefix}{Lemma}

\newcommand{\tabincell}[2]{\begin{tabular}{@{}#1@{}}#2\end{tabular}}

\newcommand{\isacodeinline}[1]{\texttt{#1}}

\usepackage{array}
\newcommand{\PreserveBackslash}[1]{\let\temp=\\#1\let\\=\temp}
\newcolumntype{C}[1]{>{\PreserveBackslash\centering}p{#1}}
\newcolumntype{R}[1]{>{\PreserveBackslash\raggedleft}p{#1}}
\newcolumntype{L}[1]{>{\PreserveBackslash\raggedright}p{#1}}

\isabellestyle{tt}

\let\endisabellecode=\endisabellecode

\newcommand{\isacodeftsz}{\footnotesize}  

\begin{document}

\title{Reasoning About Information Flow Security of Separation Kernels with Channel-based Communication}

\titlerunning{abbreviated title}  
%
\author{Yongwang Zhao\inst{1,2} \and David San\'{a}n\inst{1} \and Fuyuan Zhang\inst{1} \and Yang Liu\inst{1}}
\authorrunning{Yongwang Zhao et al.} 
%
%
\institute{School of Computer Engineering, Nanyang Technological University, Singapore
\and
School of Computer Science and Engineering, Beihang University, Beijing, China}

\maketitle              

\begin{abstract}
Assurance of information flow security by formal methods is mandated in security certification of separation kernels. 
As an industrial standard for separation kernels, ARINC 653 has been complied with by mainstream separation kernels. Security of functionalities defined in ARINC 653 is thus very important for the development and certification of separation kernels. 
This paper presents the first effort to formally specify and verify separation kernels with ARINC 653 channel-based communication. We provide a reusable formal specification and security proofs for separation kernels in Isabelle/HOL. 
During reasoning about information flow security, we find some security flaws in the ARINC 653 standard, which can cause information leakage, and fix them in our specification. We also validate the existence of the security flaws in two open-source ARINC 653 compliant separation kernels. 
\end{abstract}

\section{Introduction}
Separation kernels \cite{Rushby81} create a secure environment by providing temporal and spatial separation of applications and ensure that there are no unintended channels for information flows between partitions other than those explicitly provided. 
Separation kernels decouple the verification of applications in partitions from the verification of the kernels themselves. They are often sufficiently small and straightforward to allow formal verification of their correctness. 
Assurance of information flow security \cite{sabelfeld03} by formal methods is mandated in Separation Kernel Protection Profile (SKPP) \cite{SKPP07} and certifying separation kernels to highest Common Criteria evaluation levels (EAL 6 or 7) is always accomplished by formally verifying information flow security. 

Traditionally, security and safety of critical systems are assured and certified by using two kinds of separation kernels respectively, such as VxWorks 653 \cite{vxworks653} for safety-critical systems and VxWorks MILS \cite{vxworksmils} for security-critical systems. 
A trend in this field is to integrate safe and secure functionalities into one separation kernel. For instance, PikeOS \cite{pikeos}, LynxSecure \cite{lynxsec} and open-source XtratuM \cite{xtratum} are designed to support both safety critical and security critical solutions. 
As an industrial standard for safety-critical separation kernels, ARINC 653 \cite{ARINC653p1} aims at improving safety and certification process of safety-critical systems, which has been complied with by the mainstream separation kernels such as PikeOS, VxWorks 653 and XtratuM. 
Therefore, in order to develop ARINC 653 compliant secure separation kernels, it is necessary to assure security of the functionalities defined in ARINC 653. A security verified specification and its mechanically checked proofs of ARINC 653 are significant for the development and certification of separation kernels.

In separation kernels, Inter-Partition Communication (IPC) is a major mechanism to implement controlled information flows, but if the mechanism is not well designed, IPC can also contain covert channels \cite{millen99} to leak information between applications. ARINC 653 defines the functionalities and services of a \emph{channel-based communication} mechanism for IPC. 
Although formal specification \cite{craig07,velykis10,Verb14,Klaus15} and verification \cite{Heitmeyer08,Richards10,Wilding10,freit11,Murray13,Dam13,sanan14} of information flow security on separation kernels have been widely studied in academia and industry, information flow security of separation kernels with ARINC 653 channel-based communication has not been studied to date. To the best of our knowledge, this paper is the first effort on this topic. 

In this paper, we present a formal specification and its security proofs\footnote{The complete specification and proofs are available at http://www3.ntu.edu.sg/home/ywzhao/xkernel.htm} of separation kernels with ARINC 653 channel-based communication in Isabelle/HOL \cite{Nipkow02}. In detail, the technical contributions of this work are as follows.

\begin{enumerate}
\vspace{-7pt}
\item \label{contr:1} We provide a mechanically checked formal specification which comprises a generic execution model for separation kernels and an event specification for ARINC 653. We introduce two security domains: a \emph{scheduler} and a \emph{message transmitter}, and their security policies according to the characteristics of scheduling and IPC of separation kernels. The event specification models all IPC services defined in ARINC 653 ({\sectprefix} \ref{sec:sys_model_func_spec}). 
\item \label{contr:2} We define a set of information flow security properties and an inference framework to sketch out the implications between security properties. We provide the security proofs to indicate information flow security of the specification ({\sectprefix} \ref{sec:info_flow}). 
\item \label{contr:3} We find some security flaws, i.e., covert channels to leak information, in the ARINC 653 standard when proving our original specification that is completely compliant with ARINC 653, and fix them by a redesign of the specification. We also validate the existence of the security flaws in two open-source ARINC 653 compliant separation kernels, i.e., XtratuM and POK \cite{pok}. The cost of this work is in total 8 person-months ({\sectprefix} \ref{sec:reslt_disc}).
\end{enumerate}

\section{Challenges and Approach Overview}
\label{sec:appr_ov}
This section introduces the challenges in this work and the overview of our approach. 

\subsubsection{Challenges}
The challenges of this work are as follows.

\begin{enumerate}
\vspace{-4pt}
\item \emph{High complexity of the ARINC 653 standard}: the standard specifies the system functionality of separation kernels using more than 40 pages of informal descriptions and standardized services using more than 60 pages. As the core part for channel-based communication, the IPC takes more than 20 pages and defines a complicated communication mechanism including queuing and sampling modes, channel buffers and port control. 
\item \emph{Enormous efforts needed by formal verification of information flow security}: As a sort of hyperproperties \cite{Clarkson10}, it is difficult to automatically verify information flow security on separation kernels so far and formal verification needs an exhausting effort. 
There exist different sorts of information flow security (e.g., in \cite{rushby92,sabelfeld03,von04,Murray12}) and relationship of them on ARINC 653 separation kernels has to be clarified for security assurance and certification to reduce the verification effort. 
\vspace{-4pt}
\end{enumerate}

\subsubsection{Analysis of the Target System}
\begin{figure}[t]
\centerline{\includegraphics[width=4.0in]{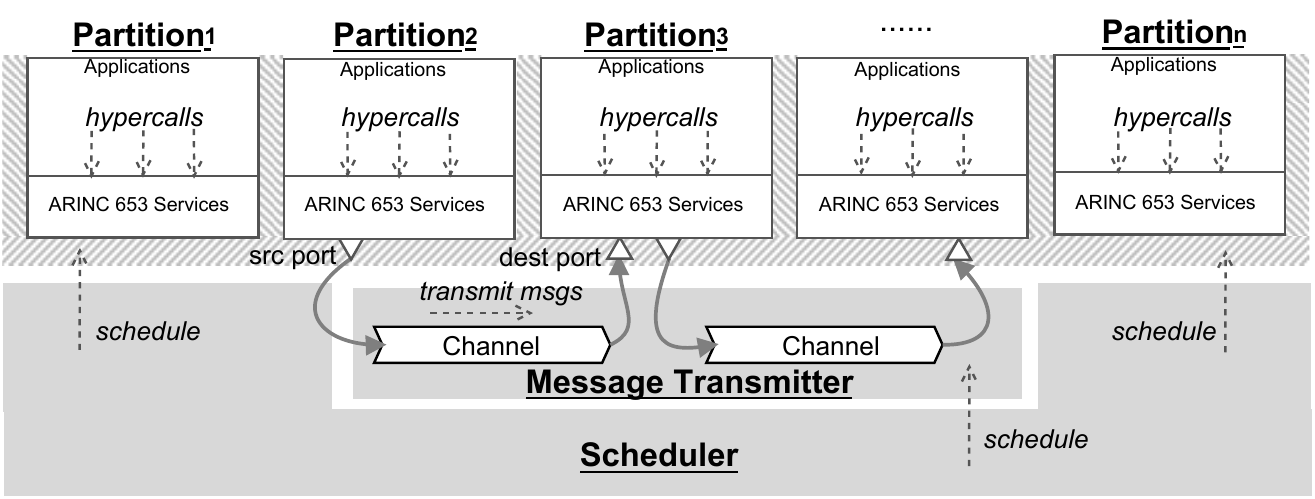}}
\caption{Architecture of the Target System}
\label{fig:arch}
\end{figure}

In order to address \emph{Challenge 1}, we are more concerned on basic functionalities of separation kernels and reduce components not related to information flow security, such as hardware interface in ARINC 653. ARINC 653 uses the inter-partition flow policy \cite{Levin07} in which communication ports and channels are associated with partitions, and all processes in a partition can access the ports configured for this partition. Moreover, some hypervisor based separation kernels, such as XtratuM, manage partitions, but processes in a partition are invisible to the kernel. Thus, we omit the concept of ``process'' and intra-partition communication between processes in ARINC 653 in the formal specification. 
The target system to be formally specified and verified is illustrated in {\figprefix} \ref{fig:arch}.

Since the latest version of ARINC 653 \cite{ARINC653p1} is targeted at single-core processing environments, our work considers single-core separation kernels and assumes there is no in-kernel concurrency as the same as in \cite{Murray13}. Many separation kernel implementations only allow blocking partitions by means of invoking a ``partition management'' hypercall, we prohibit blocking partitions in communication events.

\subsubsection{Analysis of Information Flow Security}
Traditionally, language-based information flow security \cite{sabelfeld03} handles only two-level domain: \emph{High} and \emph{Low}. The data of programs are assigned either \emph{High} or \emph{Low} labels. Security hereby means that variations of \emph{High}-level data should not cause a variation of \emph{Low}-level data. When verifying information flow security of separation kernels, the only available information is the set of configured partitions, local configurations of partitions, and the set of possible events (hypercalls) partitions can invoke. There is not any concrete information about private data of partitions. Thus, it is not possible to classify the data as \emph{High} or \emph{Low}.
Moreover, the inter-partition flow policy of ARINC 653 is an intransitive policy \cite{rushby92}, which cannot be addressed by traditional language-based information flow security. 
This problem is solved in~\cite{rushby92}, where noninterference is defined following a state-event based approach that considers intransitivity. 
In order to clarify different definitions on separation kernels, we formalize language-based information flow security in a state-event style and reason about the relationship of them. 

Traditional formulations in the state-event based approach for information flow security assume a static mapping from actions to domains, such that the domain of an action can be determined solely from the action itself \cite{rushby92}. However, in separation kernels that mapping is dynamic. When a \emph{hypercall} occurs, the kernel must consult the kernel scheduler to determine which partition is currently running, and the currently running partition is the domain of the hypercall.
In our specification, we define the \emph{scheduler} security domain for kernel scheduling, which cannot be interfered by any other domain to ensure that the scheduler security domain does not leak information via its scheduling decisions.
Since ARINC 653 only defines the channel-based communication services using ports and leaves the implementation of message transmission on channels to underlying separation kernels, we define the \emph{message transmitter} security domain, for message transmission. The transmitter also decouples message transmission from the scheduler to ensure that the scheduler is not interfered by partitions. 

\subsubsection{Analysis of the Specification and Verification Approach}
Since separation kernels usually support the deployment of partitions which are unknown in advance, it is well suited to use logical reasoning by induction for formal verification. By following the successful experiences of applying Isabelle/HOL in seL4 \cite{Murray13} and PikeOS \cite{Verb14,Klaus15}, we use Isabelle/HOL in this work.

\begin{wrapfigure}{r}{2.0in}
\vspace{-18pt}
\centerline{\includegraphics[width=2.0in]{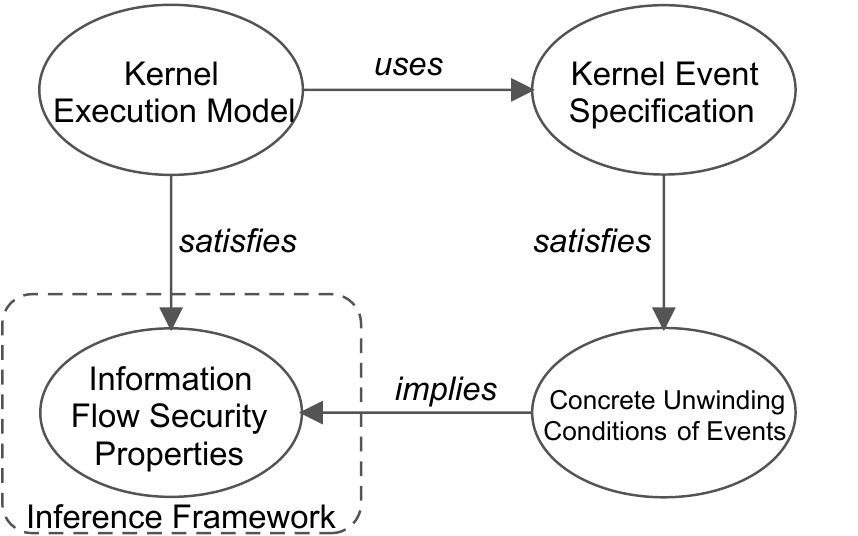}}
\caption{Verification Overview}
\label{fig:approach}
\vspace{-18pt}
\end{wrapfigure}

The verification overview of our work is briefly shown in {\figprefix} \ref{fig:approach}. 
In order to simplify the verification, we decompose the specification into two parts: an execution model for separation kernels with channel-based communication and an event specification for ARINC 653. The execution model defines basic components and a state machine of separation kernels. 
The event specification uses Isabelle/HOL functions to define the state changes when an event occurs. These concrete functions are invoked by the execution model. 
This decomposition leads to two-step proofs of information flow security. We first define a set of information flow security properties and provide an inference framework for them on the execution model. In the second step, we define a set of \emph{concrete unwinding conditions} on the concrete functions. Satisfaction of the concrete unwinding conditions implies that the events satisfy the classical unwinding conditions, and thus shows information flow security of our specification. 
The decomposition of the specification and its proofs improves their reusability for subsequent specification refinement and development of implementations, and thus reduces the verification effort.

\section{Formal Specification}
\label{sec:sys_model_func_spec}

In this section, we first introduce the kernel execution model including basic components and state-based kernel execution. Then, we present the event specification. Finally, we discuss the correctness of the formal specification. 

\subsection{Basic Components}
According to {\figprefix} \ref{fig:arch}, basic components include security domains, security policies and communication components. All these components are statically configured in ARINC 653 compliant separation kernels. 

\subsubsection{Security Domains and Policies} As illustrated in bold and underlined in {\figprefix} \ref{fig:arch}, the security domains are the scheduler, the transmitter, and the defined partitions. 
In order to discuss information flow policy, we assume a reflexive relation $\leadsto$ that specifies the allowable information flows between domains. If there is a channel from a partition \isacodeinline{a} to a partition \isacodeinline{b}, then \isacodeinline{a $\leadsto$ transmitter} and \isacodeinline{transmitter $\leadsto$ b} since we use the transmitter as the message intermediator. Since the scheduler can possibly schedule any domain, we define in the security policy that \isacodeinline{scheduler $\leadsto$ d} for any domain \isacodeinline{d}. The noninterference relation \isacodeinline{{\isasymsetminus}{\isasymleadsto}} is the complement relation of $\leadsto$ that asserts no information flow outside of $\leadsto$. 

\subsubsection{Communication Components}
As illustrated in {\figprefix} \ref{fig:arch}, IPC is conducted via messages on channels, which are defined by an abstract type \isacodeinline{Message}. Partitions have access to channels via \emph{ports} which are the endpoints of channels. A \emph{channel} links partitions and is a logical link between one source port and one or more destination ports. It also specifies the mode of transferring messages, which can be \emph{queuing} or \emph{sampling} mode. The \textbf{datatype} \isacodeinline{Channel{\isacharunderscore}Type} and \isacodeinline{Port{\isacharunderscore}Type} define these two components.

\subsubsection{System Configuration}
A significant characteristic of ARINC 653 compliant separation kernels is that partitions, policies and communication components are statically configured at built-time. 

In our specification, we use \isacodeinline{\isacommand{record}\isamarkupfalse\ Sys{\isacharunderscore}Config} to define the system configuration and \isacodeinline{\isakeyword{fixes}\ sysconf\ {\isacharcolon}{\isacharcolon}\ {\isachardoublequoteopen}Sys{\isacharunderscore}Config{\isachardoublequoteclose}} as a constant in the specification. 

\subsection{State-based Kernel Execution}
\label{subsec:exec_model}

\subsubsection{Event and State}
We consider four types of events: \emph{hypercalls}, \emph{system events}, \emph{exceptions}, and \emph{actions in partitions}. Hypercalls cover all IPC services in ARINC 653. System events are the actions of the kernel itself and include kernel initialization, scheduling and message transmission. The other two types are abstract events that can be refined in a concrete specification. Events are illustrated in {\figprefix} \ref{fig:arch} as dotted line arrows and italics. Since there is no in-kernel concurrency, all these events execute atomically. 

It is not that all events are enabled in a state. We use a function \isacodeinline{event\_enabled} to indicate whether an event can execute in a state. 
The function \isacodeinline{exec\_event} executes an event in a state and changes the state when it is enabled.
In the event specification, we define functions to implement concrete communication, scheduling and message transmission. The \isacodeinline{exec\_event} function here invokes the concrete functions.

The state is defined as \isacodeinline{\isacommand{record} State}, which consists of information about the current running partition, partition states, communication states, created ports and current value of local variables in domains. 
For a state \isacodeinline{s::State} and a sequence of events \isacodeinline{as}, \isacodeinline{execute as s} denotes the final state reached by executing \isacodeinline{as} from \isacodeinline{s}. 

\subsubsection{Domain of Events}
Events have their own execution domains. The domain of the system events is static: the domain of the event \emph{scheduling} is the scheduler; the domain of \emph{message transmission} is the transmitter. On the other hand, the domain of hypercalls is dynamic and dependent on the current state of the kernel, defined as \isacodeinline{domain{\isacharunderscore}of{\isacharunderscore}event\ s\ {\isacharparenleft}hyperc\ h{\isacharparenright}\ {\isacharequal}\ current\ s}, where \isacodeinline{current\ s} returns the currently running partition in the state \isacodeinline{s}. 

\subsubsection{State Reachability}
Since not all events are enabled in a state, some states in the type \isacodeinline{State} are not reachable from the initial state \isacodeinline{s0}. 
Let \isacodeinline{reachable s {\isasymequiv}\ {\isasymexists}as{\isachardot}\ s\ {\isacharequal}\ execute\ as\ s{\isadigit{0}}} denote that the state \isacodeinline{s} is reachable from the initial state \isacodeinline{s0}. 
According to the definition of \isacodeinline{reachable} and \isacodeinline{execute}, we have \isacodeinline{reachable\ s{\isadigit{0}}} and {\lemmaprefix} \ref{lemma:reach_s}.

\begin{lemma}
\label{lemma:reach_s}
\begin{isabellec}
{\isasymforall}s\ as{\isachardot}\ reachable\ s\ {\isasymand}\ s{\isacharprime}\ {\isacharequal}\ execute\ as\ s\ {\isasymlongrightarrow}\ reachable\ s{\isacharprime}
\end{isabellec}
\end{lemma}

\subsubsection{State Equivalence}
A key concept for information flow security is that states are \emph{identical} for a security domain. We define an equivalence relation $\sim$ \isacodeinline{d} $\sim$ on states for each domain \isacodeinline{d} such that \isacodeinline{s} $\sim$ \isacodeinline{d} $\sim$ \isacodeinline{t} if and only if states \isacodeinline{s} and \isacodeinline{t} are identical for domain \isacodeinline{d}, that is to say states \isacodeinline{s} and \isacodeinline{t} are indistinguishable for domain \isacodeinline{d}. For a set of domains \isacodeinline{D}, we define \isacodeinline{s $\approx$ D $\approx$ t $\equiv$ $\forall$d $\in$ D. s $\sim$ d $\sim$ t}.

For a partition \isacodeinline{d}, \isacodeinline{s $\sim$ d $\sim$ t} if and only if \isacodeinline{vpeq\_part s d t}, where 

\begin{isabellec}\isacodeftsz
\vspace{0.3em}
\ \ \ vpeq{\isacharunderscore}part\ s\ d\ t\ {\isasymequiv}
 vpeq{\isacharunderscore}vars\ s\ {\isacharparenleft}the\ {\isacharparenleft}{\isacharparenleft}domv\ sysconf{\isacharparenright}\ d{\isacharparenright}{\isacharparenright}\ t\ \isanewline 
\ \ \ \ \ \ \ \ \   {\isasymand}\ {\isacharparenleft}partitions\ s{\isacharparenright}\ d\ {\isacharequal}\ {\isacharparenleft}partitions\ t{\isacharparenright}\ d\ {\isasymand}\  vpeq{\isacharunderscore}part{\isacharunderscore}comm\ s\ d\ t
\vspace{0.3em}
\end{isabellec}

It means that states \isacodeinline{s} and \isacodeinline{t} are equivalent for a partition \isacodeinline{d}, when values of local variables, partition state, and communication abilities of \isacodeinline{d} on these two states are the same. 
An example of the communication ability is that if a destination queuing port \isacodeinline{p} is not empty in two states \isacodeinline{s} and \isacodeinline{t}, a partition \isacodeinline{d} has the same ability on \isacodeinline{p} in \isacodeinline{s} as in \isacodeinline{t}, because \isacodeinline{d} has the ability to receive a message from \isacodeinline{p} in these two states.
The equivalence of communication abilities defines that partition \isacodeinline{d} has the same set of ports, and that the number of messages is the same for all destination ports on states \isacodeinline{s} and \isacodeinline{t}. 

Two states \isacodeinline{s} and \isacodeinline{t} are equivalent for the \isacodeinline{scheduler} when the values of local variables of the scheduler and the current running partition on the two states are the same. The equivalence of states for the \isacodeinline{transmitter} requires that all ports, states of the ports and values of local variable are the same.

\subsection{Event Specification}
\label{sec:func_spec}

The event specification defines the concrete functions to implement the execution of events. The functionalities of separation kernels in this paper include kernel initialization, scheduling, message transmission and hypercalls. The kernel initialization considers initialization of the kernel state. Since our specification does not define processes, we only consider the partition scheduling rather than the two-level scheduling on partition and process levels in ARINC 653. Because the execution of message transmission is also under the control of scheduling, we define an abstract partition scheduling that non-deterministically chooses one partition or the transmitter as the currently executing domain. 

This subsection mainly discusses channel-based communication services in ARINC 653 and the message transmission. All events and their descriptions in the event specification are shown in {\tableprefix} \ref{tbl:events}.

\begin{table}[t]
	\centering
	\scriptsize
	\caption{Events in Our Specification} 
	\begin{tabular} {|l|l|p{8.0cm}|}
		\hline
		\textbf{No}. & \textbf{Name} & \textbf{Description of Event Specification} 
		\\
		\hline
		\multicolumn{3}{|l|}{\textbf{Hypercalls}}
		\\
		\hline
		(1) & Create\_Sampling\_Port & Create a sampling port. An identifier is assigned by the kernel and returned. 
		\\
		\hline
		(2) & Write\_Sampling\_Message & Write a message in the specified sampling port. The message overwrites the previous one.
		\\
		\hline
		(3) & Read\_Sampling\_Message & Read a message from the specified sampling port. 
		\\
		\hline
		(4) & Get\_Sampling\_Portid & Return the sampling port identifier that corresponds to a sampling port name.
		\\
		\hline
		(5) & Get\_Sampling\_Portstatus & Return the current status of the specified sampling
		port.
		\\
		\hline
		(6) & Create\_Queuing\_Port & Create a queuing port. An identifier is assigned by the kernel and returned. 
		\\
		\hline
		(7) & Send\_Queuing\_Message & Send a message in the specified queuing port. If there is sufficient space in the queuing port to accept the message, the message is inserted into the port buffer. If there is insufficient space, the message is lost.
		\\
		\hline
		(8) & Receive\_Queuing\_Message & Receive a message from the specified queuing port. If the queuing port is not empty, a message in the port buffer is removed and returned. If the queuing port is empty, \isacodeinline{None} is returned.
		\\
		\hline
		(9) & Get\_Queuing\_Portid & Return the queuing port identifier that corresponds to a queuing port name.
		\\
		\hline
		(10) & Get\_Queuing\_Portstatus & Return the current status of the specified queuing port.
		\\
		\hline
		(11) & Clear\_Queuing\_Port & Discard any messages in the message buffer of the specified destination port.
		\\
		\hline
		\multicolumn{3}{|l|}{\textbf{System events}}
		\\
		\hline
		(12) & Schedule & Set one partition or the transmitter as the currently running domain. 
		\\
		\hline
		(13) & Transfer\_Sampling\_Message & Copy the message in the source sampling port to all destination sampling ports of a sampling channel, if all ports of this channel have been created. 
		\\
		\hline
		(14) & Transfer\_Queuing\_Message & Copy a message in the source queuing port to the destination queuing port of a queuing channel and remove the message from the source port, if the two ports of this channel have been created and the source port is not empty. If the destination port is full, the message is lost. 
		\\
		\hline
		(15) & Init & Initialize the kernel state using the system configuration. 
		\\
		\hline
	\end{tabular}
	\label{tbl:events}
	\vspace{-12pt}
\end{table}

\subsubsection{Channel-based Communication Services}

\begin{figure}[t]
	\centerline{\includegraphics[width=3.0in]{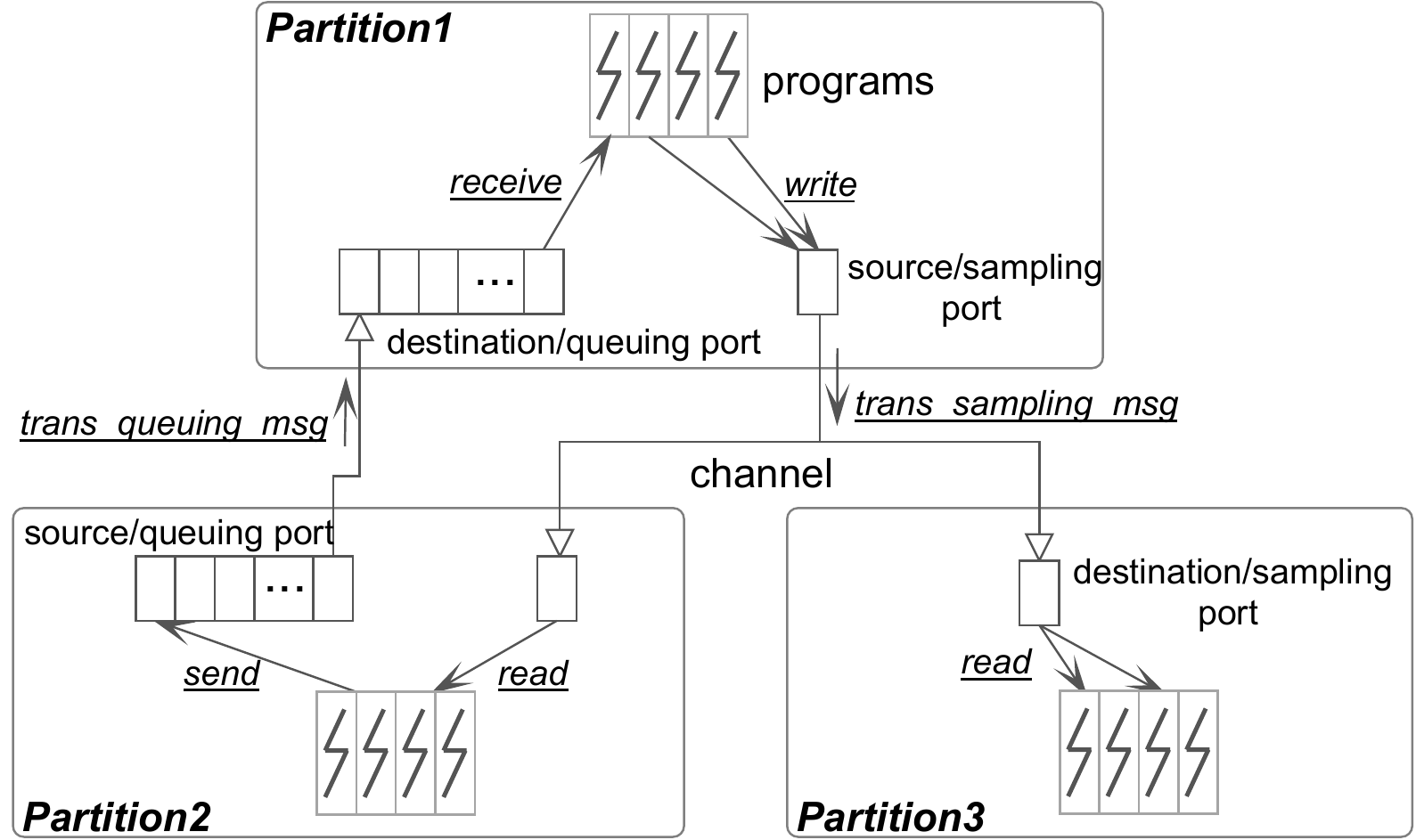}}
	\caption{Channel-based Communication in ARINC 653}
	\label{fig:comm}
	\vspace{-8pt}
\end{figure}

ARINC 653 specifies the behavior of ports and the communication services via ports in detail. Programs in a partition could use IPC by invoking these services. ARINC 653 defines eleven services for sampling and queuing ports (No. 1 $\sim$ 11 in {\tableprefix} \ref{tbl:events}). The communication architecture is illustrated in {\figprefix} \ref{fig:comm}.

In the first stage of this work, we design the event specification completely based on the service behavior specified in ARINC 653. When proving the unwinding conditions on these events, we find covert channels ({\sectprefix} \ref{sec:reslt_disc} in detail) and change the service specification defined in ARINC 653 to avoid these covert channels. 
McCullough \cite{McCull88} provides three ways to avoid covert channels: unbounded buffer, process blocking and message loss. According to the discussion in {\sectprefix} \ref{sec:appr_ov}, we do not allow partition blocking in communication services. Because unbounded buffer would lead to a bigger problem of denial of service (DoS), the feasible way for our specification is to allow message loss. In order to avoid covert channels, we allow message loss when sending a message to a queuing port and transmitting a message in a queuing channel.

We use a set of functions to implement one service. For instance, the \isacodeinline{Send\_Queuing\_Message} service is implemented by function \isacodeinline{send\_queuing\_message\_maylost} as follows and a set of related functions invoked by this function. 

\begin{isabellec}\isacodeftsz
\vspace{0.3em}
\isacommand{definition}\isamarkupfalse%
\ send{\isacharunderscore}queuing{\isacharunderscore}message{\isacharunderscore}maylost\ {\isacharcolon}{\isacharcolon}\ {\isachardoublequoteopen}Sys{\isacharunderscore}Config\ {\isasymRightarrow}\ State\ {\isasymRightarrow}\ port{\isacharunderscore}id\ {\isasymRightarrow}\ Message\ {\isasymRightarrow}\ {\isacharparenleft}State\ {\isasymtimes}\ bool{\isacharparenright}{\isachardoublequoteclose}\ \isakeyword{where}\isanewline
\ \ {\isachardoublequoteopen}send{\isacharunderscore}queuing{\isacharunderscore}message{\isacharunderscore}maylost\ sc\ s\ p\ m\ {\isasymequiv}\ \isanewline
\ \ \ \ \ {\isacharparenleft}if{\isacharparenleft}{\isasymnot}\ is{\isacharunderscore}a{\isacharunderscore}queuingport\ s\ p\  {\isasymor}\ {\isasymnot}\ is{\isacharunderscore}source{\isacharunderscore}port\ s\ p \isanewline
\ \ \ \ \ \ \ \ \ \ \ \ \ \ \ \ {\isasymor}\ {\isasymnot}\ is{\isacharunderscore}a{\isacharunderscore}port{\isacharunderscore}of{\isacharunderscore}partition\ s\ p {\isacharparenright} \ then \ {\isacharparenleft}s{\isacharcomma}\ False{\isacharparenright}\isanewline
\ \ \ \ \ \ else\ if\ is{\isacharunderscore}full{\isacharunderscore}portqueuing\ sc\ s\ p\ then \ {\isacharparenleft}s{\isacharcomma}\ True{\isacharparenright}\isanewline
\ \ \ \ \ \ else \ {\isacharparenleft}insert{\isacharunderscore}msg{\isadigit{2}}queuing{\isacharunderscore}port\ s\ p\ m{\isacharcomma}\ True{\isacharparenright}{\isacharparenright}{\isachardoublequoteclose}%
\vspace{0.3em}
\end{isabellec}

As specified in the \isacodeinline{Send\_Queuing\_Message} service in ARINC 653, when sending a message via a queuing port, it fails if either the specified port does not exist, or it is not a source port, or it is not in current partition. When the port is full, the calling process is blocked. Since blocking is not considered in this paper, we just discard the message. 

\subsubsection{Message Transmission on Channels}
ARINC 653 does not define the functionalities of message transmission and leaves its implementation to underlying separation kernels. We design a basic specification of the message transmission in this paper.

The message transmission on channels is shown in {\figprefix} \ref{fig:comm}. ARINC 653 has two modes of channel-based communication: sampling and queuing mode. The multicast message that is sent from a single source to more than one destination is supported in sampling mode. The queuing mode only supports the unicast message. 
In sampling mode, a message transmission on a channel copies the message in the source sampling port of the channel to the buffers of all destination sampling ports of the channel. Whilst in queuing mode, a message transmission on a channel copies a message in the buffer of the source queuing port, removes it from this buffer and stores the message into the buffer of the destination queuing port of the channel. When the buffer of the destination queuing port is full, the message is discarded.

For instance, the message transmission in queuing mode is defined as follows. If the source and destination port have been created and there are messages in the buffer of the source port, a message in the buffer is removed and inserted into the buffer of the destination port. When the buffer of the destination port is full, the message is discarded. 

\begin{isabellec}\isacodeftsz
\vspace{0.3em}
\isacommand{primrec}\isamarkupfalse%
\ transf{\isacharunderscore}queuing{\isacharunderscore}msg{\isacharunderscore}maylost\ {\isacharcolon}{\isacharcolon}\ {\isachardoublequoteopen}Sys{\isacharunderscore}Config\ {\isasymRightarrow}\ State\ {\isasymRightarrow}\ Channel{\isacharunderscore}Type \isanewline
{\isasymRightarrow}\ State{\isachardoublequoteclose}\ \isakeyword{where}\isanewline
\ \ {\isachardoublequoteopen}transf{\isacharunderscore}queuing{\isacharunderscore}msg{\isacharunderscore}maylost\ sc\ s\ {\isacharparenleft}Channel{\isacharunderscore}Queuing\ {\isacharunderscore}\ sn\ dn{\isacharparenright}\ {\isacharequal}\ \isanewline
\ \ \ \ {\isacharparenleft}let\ sp\ {\isacharequal}\ get{\isacharunderscore}portid{\isacharunderscore}by{\isacharunderscore}name\ s\ sn{\isacharsemicolon} dp\ {\isacharequal}\ get{\isacharunderscore}portid{\isacharunderscore}by{\isacharunderscore}name\ s\ dn\ in\isanewline
\ \ \ \ \ \ if\ sp\ {\isasymnoteq}\ None\ {\isasymand}\ dp\ {\isasymnoteq}\ None\ {\isasymand}\ has{\isacharunderscore}msg{\isacharunderscore}inportqueuing\ s\ {\isacharparenleft}the\ sp{\isacharparenright}\ then\isanewline
\ \ \ \ \ \ \ \ let\ sm\ {\isacharequal}\ remove{\isacharunderscore}msg{\isacharunderscore}from{\isacharunderscore}queuingport\ s\ {\isacharparenleft}the\ sp{\isacharparenright}\ in\isanewline
\ \ \ \ \ \ \ \ \ \ \ \ if\ is{\isacharunderscore}full{\isacharunderscore}portqueuing\ sc\ {\isacharparenleft}fst\ sm{\isacharparenright}\ {\isacharparenleft}the\ dp{\isacharparenright}\ then s \isanewline
\ \ \ \ \ \ \ \ \ \ \ \ else\isanewline
\ \ \ \ \ \ \ \ \ \ \ \ \ \ insert{\isacharunderscore}msg{\isadigit{2}}queuing{\isacharunderscore}port\ {\isacharparenleft}fst\ sm{\isacharparenright}\ {\isacharparenleft}the\ dp{\isacharparenright}\ {\isacharparenleft}the\ {\isacharparenleft}snd\ sm{\isacharparenright}{\isacharparenright}\ \isanewline
\ \ \ \ \ \ else\ s {\isacharparenright}{\isachardoublequoteclose}\ {\isacharbar}\isanewline
\ \ {\isachardoublequoteopen}transf{\isacharunderscore}queuing{\isacharunderscore}msg{\isacharunderscore}maylost\ sc\ s\ {\isacharparenleft}Channel{\isacharunderscore}Sampling\ {\isacharunderscore}\ {\isacharunderscore}\ {\isacharunderscore}{\isacharparenright}\ {\isacharequal}\ s{\isachardoublequoteclose}
\vspace{0.3em}
\end{isabellec}

\subsection{Correctness of Formal Specification}
To assure the correctness of our specification, beside the manual validation by inspecting the Isabelle/HOL specification, we prove that functionalities of the specified services are correct w.r.t. the ARINC 653 informal description~\cite{ARINC653p1} by means of 33 lemmas for events and invariants. Due to the atomicity of event execution, the correctness of an event can be specified and proved by pre- and post-conditions of the event in Hoare logic \cite{Hoare69}, i.e., $\{P\} \ C \ \{Q\}$, where $C$ is the Isabelle function implementing the event, $P$ and $Q$ are the pre- and post-conditions respectively. Since the execution of events always terminate, our specification is a total correctness specification. Termination is ensured by using the \textbf{primrec} and \textbf{definition} in Isabelle/HOL to define the functions in our specification and proved automatically in Isabelle/HOL. 
For instance, the correctness lemma for the event \isacodeinline{Create\_Sampling\_Port} is as follows. 
The pre-condition is that the port named \isacodeinline{p} is configured, has not been created and is a port of the currently running partition. Under the pre-condition, the execution of \isacodeinline{create{\isacharunderscore}sampling {\isacharunderscore}port} returns a pair of the new state and the assigned identifier of the created port. The post-condition ensures that the identifier (\isacodeinline{the (snd r)}) is stored in the \isacodeinline{ports} in the new state (\isacodeinline{ports (comm (fst r))}).

\begin{lemma}[Correctness of Create\_Sampling\_Port]
\label{lemma:cor_wsm}
\begin{isabellec}
\{ get{\isacharunderscore}samplingport{\isacharunderscore}conf\ sysconf\ p\ {\isasymnoteq}\ None {\isasymand}
get{\isacharunderscore}portid{\isacharunderscore}by{\isacharunderscore}name\ s\ p\ {\isacharequal}\ None {\isasymand}
p\ {\isasymin}\ get{\isacharunderscore}partition{\isacharunderscore}cfg{\isacharunderscore}ports{\isacharunderscore}byid\ sysconf\ {\isacharparenleft}current\ s{\isacharparenright} \}  \isanewline
r\ {\isacharequal}\ create{\isacharunderscore}sampling{\isacharunderscore}port\ sysconf\ s\ p \isanewline
\{
{\isacharparenleft}ports\ {\isacharparenleft}comm\ {\isacharparenleft}fst\ r{\isacharparenright}{\isacharparenright}{\isacharparenright}\ {\isacharparenleft}the\ {\isacharparenleft}snd\ r{\isacharparenright}{\isacharparenright}\ {\isasymnoteq}\ None
\}
\end{isabellec}
\end{lemma}

Functional correctness requires to prove invariants on the data structures defining the state. An invariant is a safety property and defined on states as a predicate $\psi$ \isacodeinline{s}. It is preserved in all reachable states by proving the invariant theorem: \isacodeinline{reachable\ s\ {\isasymLongrightarrow}\ $\psi$\ s}. 
A typical invariant is the predicate \isacodeinline{port\_consistent s}. We use a set to store created ports. The port state (e.g., the messages currently in the port) is defined as \isacodeinline{Ports\ {\isacharequal}\ {\isachardoublequoteopen}port{\isacharunderscore}id\ {\isasymrightharpoonup}\ Port{\isacharunderscore}Type{\isachardoublequoteclose}}. Ports belong to different partitions that is defined as \isacodeinline{part{\isacharunderscore}ports\ {\isacharcolon}{\isacharcolon}\ {\isachardoublequoteopen}port{\isacharunderscore}id\ {\isasymrightharpoonup}\ partition{\isacharunderscore}id{\isachardoublequoteclose}}. The \isacodeinline{port\_consistent s} requires that the created port set and the domains of these two partial functions are the same in any reachable states. The invariant theorem is proved by {\lemmaprefix} \ref{lemma:reach_s} and other two lemmas: (1) \isacodeinline{$\psi$\ s{\isadigit{0}}} and (2) \isacodeinline{{\isasymforall}s\ as{\isachardot}\ $\psi$\ s\ {\isasymand}\ s{\isacharprime}\ {\isacharequal}\ execute\ as\ s\ {\isasymlongrightarrow}\ $\psi$\ s{\isacharprime}}.

\section{Information Flow Security and Proofs}
\label{sec:info_flow}
This section first presents a set of information flow security properties defined on the execution model, which includes the original definitions of noninterference \cite{rushby92}, nonleakage \cite{von04} and noninfluence \cite{von04}, and their variants. Nonleakage is language-based information flow security and noninfluence is the combination of noninterference and nonleakage. 
Then, we present an overview of our proof structure and the proofs which include an inference framework of these properties and the security proofs of our event specification.

\subsection{Formalizing Noninterference}
Since intransitive policies could be used to specify channel control policies \cite{rushby92}, we consider intransitive noninterference in this paper. 
The essence of noninterference on separation kernels is that a partition \isacodeinline{d} cannot distinguish the final states between executing a sequence of events \isacodeinline{as} and executing its purged sequence from the initial state. In the purged sequence, the events of partitions that are not allowed to pass information to \isacodeinline{d} directly or indirectly are removed.

In order to express the allowed information flow for intransitive policies, we employ the function \isacodeinline{sources} \cite{rushby92}, which takes a sequence of events \isacodeinline{as} and a target domain \isacodeinline{d} and yields the set of domains that are allowed to pass information to \isacodeinline{d} when \isacodeinline{as} occurs. Due to the dependency of event domains on states, the \isacodeinline{sources} function in our specification depends on the current state \isacodeinline{s}. 
The \isacodeinline{sources} function is used to define the classical purge function, \isacodeinline{ipurge}, in terms of which security properties are formulated. The \isacodeinline{ipurge as s d} yields the sequence of events \isacodeinline{as}, where all events that are not allowed to pass information to \isacodeinline{d} directly or indirectly when \isacodeinline{as} is executed from \isacodeinline{s} are removed.

We use the abbreviation \isacodeinline{s $\triangleleft$ as $\cong$ t $\triangleleft$ bs @ d} for the observational equivalence. It denotes that \isacodeinline{d} is identical in the two final states after executing \isacodeinline{as} from \isacodeinline{s} (by \isacodeinline{execute\ as\ s}) and executing \isacodeinline{bs} from \isacodeinline{t}. 
Traditionally, this equivalence is defined using a projection function \isacodeinline{output} which returns the observed results on a state by a domain. In this paper, we have combined the \isacodeinline{output} in the state equivalence presented in {\subsectprefix} \ref{subsec:exec_model}. This allows us to avoid the unwinding condition of \emph{output consistent}. 
We define the classical nontransitive noninterference \cite{rushby92} on our execution model as follows.

\begin{isabellec}
\vspace{0.3em}
\ \ \ noninterference\ {\isasymequiv}\ {\isasymforall}d\ as{\isachardot}\ {\isacharparenleft}s{\isadigit{0}}\ {\isasymlhd}\ as\ {\isasymcong}\ s{\isadigit{0}}\ {\isasymlhd}\ {\isacharparenleft}ipurge\ as\ s{\isadigit{0}}\ d{\isacharparenright}\ {\isacharat}\ d{\isacharparenright}
\vspace{0.3em}
\end{isabellec}

In the definition of noninterference, the \isacodeinline{ipurge} function only deletes all unsuitable events. A strong version of noninterference is introduced in \cite{von04} to handle arbitrary insertion and deletion of secret events. Oheimb \cite{von04} says that the strong noninterference and the original one are equivalent in deterministic cases. We define this strong version of noninterference on the execution model as \isacodeinline{weak{\isacharunderscore}noninterference}, since \isacodeinline{noninterference} implies \isacodeinline{weak\_noninterference} on our execution model.

The above definitions of noninterference are based on the initial state \isacodeinline{s0}, but separation kernels usually support \emph{warm} or \emph{cold start} and they may start to execute from a non-initial state. Therefore, we define a more general version of noninterference as follows based on the \isacodeinline{reachable} function. This general noninterference requires that the system starting from any reachable state is secure. It is obvious that this noninterference implies the classical noninterference due to the lemma: \isacodeinline{reachable s0}. 

\begin{isabellec}
\vspace{0.3em}
\ \ \ noninterference{\isacharunderscore}r\ {\isasymequiv}\ {\isasymforall}d\ as\ s{\isachardot}\ reachable\ s\ {\isasymlongrightarrow} \isanewline
\ \ \ \ \ \ \ \ \ \ \ \ \ \ \ \ \ \ \ \ \ \ \ {\isacharparenleft}s\ {\isasymlhd}\ as\ {\isasymcong}\ s\ {\isasymlhd} 
{\isacharparenleft}ipurge\ as\ s\ d{\isacharparenright}\ {\isacharat}\ d{\isacharparenright}
\vspace{0.3em}
\end{isabellec}

\subsection{Formalizing Nonleakage and Noninfluence}
Language-based information flow security is generalized to arbitrary multi-domain policies in \cite{von04} as a new notion \emph{nonleakage}. 
\emph{Nonleakage} and \emph{noninterference} are also combined in \cite{von04} as a new notion \emph{noninfluence}. 
Murray et al. \cite{Murray12} have extended the original definition of nonleakage and noninfluence and defined the general forms of them for operating systems based on the scheduler. We use Murray's definitions and define them on our execution model as follows. 

\begin{isabellec}
\vspace{0.3em}
\ \ \ nonleakage\ {\isasymequiv}\ {\isasymforall}d\ as\ s\ t{\isachardot}\ reachable\ s\ {\isasymand}\ reachable\ t\ {\isasymlongrightarrow}\isanewline
\ \ \ \ \ \ \ \ \ \ \   {\isacharparenleft}s\ {\isasymsim}\ {\isacharparenleft}scheduler\ sysconf{\isacharparenright}\ {\isasymsim}\ t{\isacharparenright}\ {\isasymlongrightarrow}\ {\isacharparenleft}s\ {\isasymapprox}\ {\isacharparenleft}sources\ as\ s\ d{\isacharparenright}\ {\isasymapprox}\ t{\isacharparenright}\isanewline
\ \ \ \ \ \ \ \ \ \ \   {\isasymlongrightarrow}\ {\isacharparenleft}s\ {\isasymlhd}\ as\ {\isasymcong}\ t\ {\isasymlhd}\ as\ {\isacharat}\ d{\isacharparenright}
\vspace{0.3em}
\end{isabellec}

\begin{isabellec}
\vspace{0.3em}
\ \ \ noninfluence\ {\isasymequiv}\ {\isasymforall}\ d\ as\ bs\ s\ t\ {\isachardot}\ reachable\ s\ {\isasymand}\ reachable\ t\ {\isasymlongrightarrow} \isanewline 
\ \ \ \ \ \ {\isacharparenleft}s\ {\isasymapprox}\ {\isacharparenleft}sources\ as\ s\ d{\isacharparenright}\ {\isasymapprox}\ t{\isacharparenright}\ {\isasymlongrightarrow}\ {\isacharparenleft}s\ {\isasymsim}\ {\isacharparenleft}scheduler\ sysconf{\isacharparenright}\ {\isasymsim}\ t{\isacharparenright}\ {\isasymlongrightarrow}\isanewline 
\ \ \ \ \ \ ipurge\ as\ s\ d\ {\isacharequal}\ ipurge\ bs\ s\ d\ {\isasymlongrightarrow}\ {\isacharparenleft}s\ {\isasymlhd}\ as\ {\isasymcong}\ t\ {\isasymlhd}\ bs\ {\isacharat}\ d{\isacharparenright}
\vspace{0.3em}
\end{isabellec}

The intuitive meaning of nonleakage is that if the secret data is not leaked initially, the secret data should not be leaked during executing a sequence of events. 
Separation kernels are said to preserve \emph{nonleakage} when for any pair of reachable states \isacodeinline{s} and \isacodeinline{t} and observing domain \isacodeinline{d}, if (1) \isacodeinline{s} and \isacodeinline{t} are equivalent for all domains that may (directly or indirectly) interfere with \isacodeinline{d} during the run of \isacodeinline{as}, i.e., \isacodeinline{s\ {\isasymapprox}\ {\isacharparenleft}sources\ as\ s\ d{\isacharparenright}\ {\isasymapprox}\ t}, and (2) the same domain is currently running in both states, i.e., \isacodeinline{s\ {\isasymsim}\ {\isacharparenleft}scheduler\ sysconf{\isacharparenright}\ {\isasymsim}\ t}, then \isacodeinline{s} and \isacodeinline{t} are observationally equivalent for \isacodeinline{d} when executing \isacodeinline{as}. 
Murray's definition of noninfluence is a weak one, we propose a strong one according to the Oheimb's noninfluence by extending the scheduler and state reachability as follows. 

\begin{isabellec}
\vspace{0.3em}
\ \ \ strong{\isacharunderscore}noninfluence\ {\isasymequiv}\ {\isasymforall}\ d\ as\ s\ t\ {\isachardot}\ reachable\ s\ {\isasymand}\ reachable\ t\ {\isasymlongrightarrow} \isanewline
\ \ \ \ \ \ \ \ \ \ \  {\isacharparenleft}s\ {\isasymapprox}\ {\isacharparenleft}sources\ as\ s\ d{\isacharparenright}\ {\isasymapprox}\ t{\isacharparenright}\  {\isasymlongrightarrow}\ {\isacharparenleft}s\ {\isasymsim}\ {\isacharparenleft}scheduler\ sysconf{\isacharparenright}\  {\isasymsim}\ t{\isacharparenright}\isanewline
\ \ \ \ \ \ \ \ \ \ \  {\isasymlongrightarrow}\ {\isacharparenleft}s\ {\isasymlhd}\ as\ {\isasymcong}\ t\ {\isasymlhd}\ {\isacharparenleft}ipurge\ as\ t\ d{\isacharparenright}\ {\isacharat}\ d{\isacharparenright}
\vspace{0.3em}
\end{isabellec}

\subsection{Proof Structure}
\begin{figure}[t]
\centerline{\includegraphics[width=4.2in]{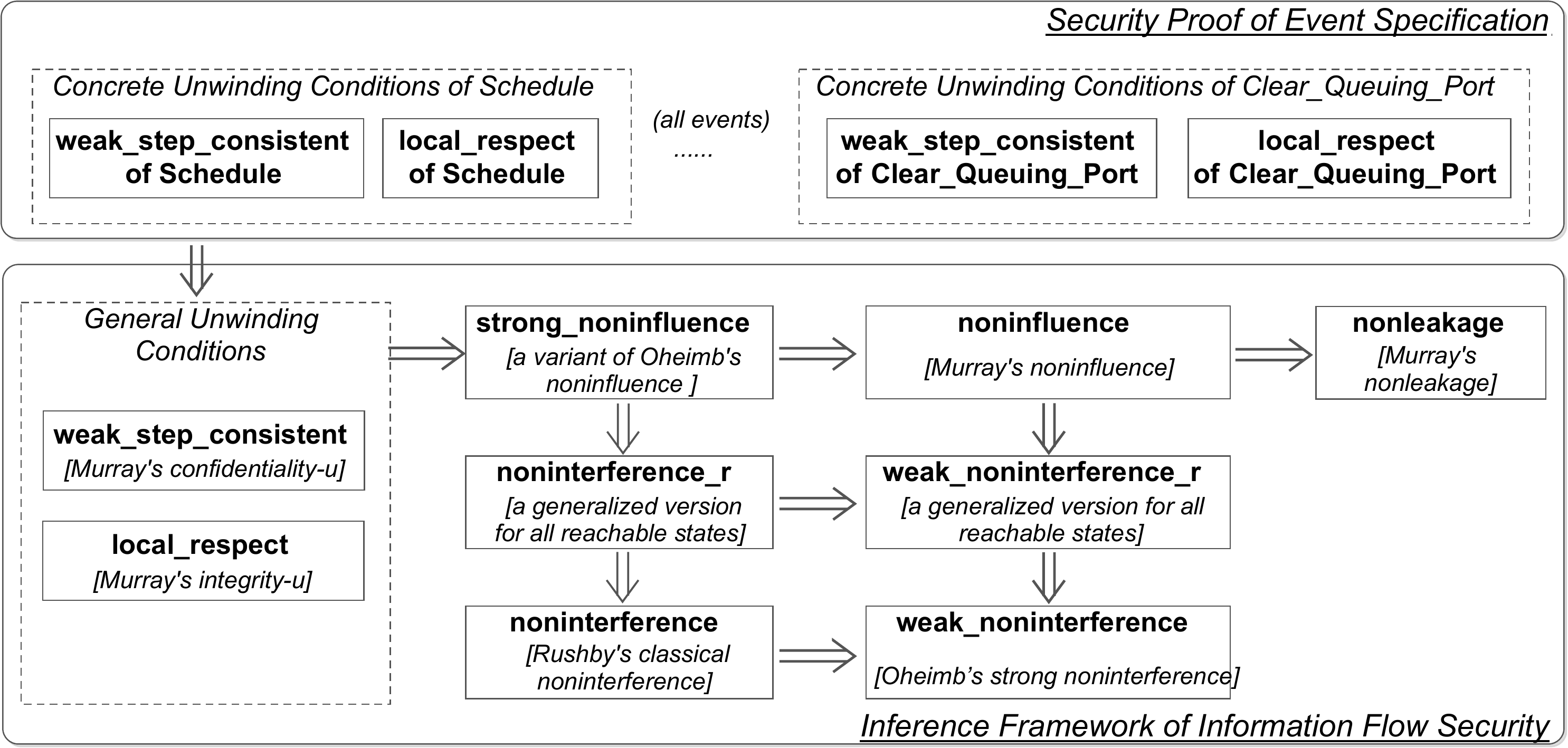}}
\caption{Proof Structure}
\label{fig:proof_struct}
\vspace{-12pt}
\end{figure}

As discussed in {\sectprefix} \ref{sec:appr_ov}, proofs of information flow security on our specification comprise two parts: an inference framework of information flow security properties on the execution model and security proofs of the event specification. The proof structure of this work is shown in {\figprefix} \ref{fig:proof_struct}, where an arrow means the implication between properties. In the next two subsections, we discuss the two parts of proofs in turn. 

\subsection{Inference Framework of Information Flow Security}
In order to clarify different properties of information flow security on our specification, we provide an inference framework on the execution model as shown in the lower part of {\figprefix} \ref{fig:proof_struct}. We have proven all implication relations between these properties on the execution model. We could see that the property \isacodeinline{strong\_noninfluence} is the strongest one and if this property is satisfied, so are all other properties.

The standard proof of information flow security properties is discharged by proving a set of unwinding conditions \cite{rushby92} that examine individual execution steps of the system. Our work follows this approach. 
In order to prove \isacodeinline{strong\_noninfluence}, we define two general unwinding conditions, \isacodeinline{weak\_step\_consistent} and \isacodeinline{local\_respect}, as follows. As there is no \isacodeinline{output} function in our specification, we do not define the classical unwinding condition of \emph{output consistent}. 

\begin{isabellec}
\vspace{0.3em}
\ \ \ weak{\isacharunderscore}step{\isacharunderscore}consistent\ {\isasymequiv}\ {\isasymforall}\ d\ a\ s\ t\ {\isachardot}\ reachable\ s\ {\isasymand}\ reachable\ t\ {\isasymlongrightarrow}\isanewline
\ \ \ \ \ \ \ \ \  
 {\isacharparenleft}s\ {\isasymsim}\ d\ {\isasymsim}\ t{\isacharparenright}\ {\isasymand}\ {\isacharparenleft}s\ {\isasymsim}\ {\isacharparenleft}scheduler\ sysconf{\isacharparenright}\ {\isasymsim}\ t{\isacharparenright}\ {\isasymand}\ \isanewline
\ \ \ \ \ \ \ \ \ {\isacharparenleft}{\isacharparenleft}domain{\isacharunderscore}of{\isacharunderscore}event\ s\ a{\isacharparenright}\ {\isasymleadsto}\ d{\isacharparenright}\ {\isasymand}\ {\isacharparenleft}s\ {\isasymsim}\ {\isacharparenleft}domain{\isacharunderscore}of{\isacharunderscore}event\ s\ a{\isacharparenright}\ {\isasymsim}\ t{\isacharparenright} \isanewline
\ \ \ \ \ \ \ \ \ {\isasymlongrightarrow}\ {\isacharparenleft}{\isacharparenleft}exec{\isacharunderscore}event\ s\ a{\isacharparenright}\ {\isasymsim}\ d\ {\isasymsim}\ {\isacharparenleft}exec{\isacharunderscore}event\ t\ a{\isacharparenright}{\isacharparenright}
\vspace{0.3em}
\end{isabellec}

\begin{isabellec}
\vspace{0.3em}
\ \ \ local{\isacharunderscore}respect\ {\isasymequiv}\ {\isasymforall}\ a\ d\ s\ s{\isacharprime}{\isachardot}\ reachable\ s\ {\isasymlongrightarrow}\isanewline
\ \ \  {\isacharparenleft}{\isacharparenleft}domain{\isacharunderscore}of{\isacharunderscore}event\ s\ a{\isacharparenright}\ {\isasymsetminus}{\isasymleadsto}\ d{\isacharparenright}\ {\isasymand}\ {\isacharparenleft}s{\isacharprime}\ {\isacharequal}\ exec{\isacharunderscore}event\ s\ a{\isacharparenright}\ {\isasymlongrightarrow}\ {\isacharparenleft}s\ {\isasymsim}\ d\ {\isasymsim}\ s{\isacharprime}{\isacharparenright}
\vspace{0.3em}
\end{isabellec}

The \isacodeinline{weak\_step\_consistent} means that for any pair of reachable states \isacodeinline{s} and \isacodeinline{t}, and any observing domain \isacodeinline{d}, the next states after executing any event \isacodeinline{a} on \isacodeinline{s} and \isacodeinline{t} are indistinguishable for \isacodeinline{d}, i.e., \isacodeinline{{\isacharparenleft}exec{\isacharunderscore}event\ s\ a{\isacharparenright}\ {\isasymsim}\ d\ {\isasymsim}\ {\isacharparenleft}exec{\isacharunderscore}event\ t\ a{\isacharparenright}}, if \isacodeinline{s} and \isacodeinline{t} are indistinguishable for \isacodeinline{d}, the same domain is currently running in \isacodeinline{s} and \isacodeinline{t}, the domain of event \isacodeinline{a} in state \isacodeinline{s} can interference with \isacodeinline{d}, and \isacodeinline{s} and \isacodeinline{t} are indistinguishable for the domain of event \isacodeinline{a}.
The \isacodeinline{weak\_step\_consistent} is the same as \isacodeinline{confidentiality{-\allowbreak}u} proposed in \cite{Murray12}. 
The \isacodeinline{local\_respect} is the same as \isacodeinline{integrity{-\allowbreak}u} in \cite{Murray12}, which means that an event \isacodeinline{a} that executes in some state \isacodeinline{s} can affect only those domains to which the domain executing event \isacodeinline{a} is allowed to send information.

\subsection{Security Proofs of Event Specification}
The second step of proofs is to show security of the event specification. 
From definitions of the two general unwinding conditions, we could see that in order to prove the satisfaction of the two conditions on our specification, we can induct on each type of events in separation kernels and prove that each concrete event satisfies the two conditions. 
Therefore, we define a set of \emph{concrete unwinding conditions} for all events. Satisfaction of the concrete unwinding conditions of one event implies that the event satisfies the general unwinding conditions. For instance, {\lemmaprefix} \ref{thm:crtqueport_lcrsp} and \ref{thm:crtqueport_wstpc} show the concrete unwinding conditions for event \isacodeinline{Create\_Queuing\_Port}.

\begin{lemma}[Local\_respect of creating\_queuing\_port]
\label{thm:crtqueport_lcrsp}
\begin{isabellec}
reachable\ s {\isasymand} 
is{\isacharunderscore}a{\isacharunderscore}partition\ sysconf\ {\isacharparenleft}current\ s{\isacharparenright} {\isasymand}
{\isacharparenleft}current\ s{\isacharparenright}\ {\isasymsetminus}{\isasymleadsto}\ d {\isasymand} \isanewline
s{\isacharprime}\ {\isacharequal}\ fst\ {\isacharparenleft}create{\isacharunderscore}queuing{\isacharunderscore}port\ sysconf\ s\ pname{\isacharparenright} 
{\isasymLongrightarrow}
s\ {\isasymsim}\ d\ {\isasymsim}\ s{\isacharprime}
\end{isabellec}
\end{lemma}

\begin{lemma}[Weak\_step\_consistent of creating\_queuing\_port]
\label{thm:crtqueport_wstpc}
\begin{isabellec}
is{\isacharunderscore}a{\isacharunderscore}partition\ sysconf\ {\isacharparenleft}current\ s{\isacharparenright} {\isasymand} 
reachable\ s\ {\isasymand}\ reachable\ t {\isasymand}  \isanewline
s\ {\isasymsim}\ d\ {\isasymsim}\ t {\isasymand} 
s\ {\isasymsim}\ {\isacharparenleft}scheduler\ sysconf{\isacharparenright}\ {\isasymsim}\ t {\isasymand} 
{\isacharparenleft}current\ s{\isacharparenright}\ {\isasymleadsto}\ d {\isasymand} \isanewline
s\ {\isasymsim}\ {\isacharparenleft}current\ s{\isacharparenright}\ {\isasymsim}\ t {\isasymand} 
s{\isacharprime}\ {\isacharequal}\ fst\ {\isacharparenleft}create{\isacharunderscore}queuing{\isacharunderscore}port\ sysconf\ s\ pname{\isacharparenright} {\isasymand} \isanewline
t{\isacharprime}\ {\isacharequal}\ fst\ {\isacharparenleft}create{\isacharunderscore}queuing{\isacharunderscore}port\ sysconf\ t\ pname{\isacharparenright} 
{\isasymLongrightarrow}
s{\isacharprime}\ {\isasymsim}\ d\ {\isasymsim}\ t{\isacharprime}
\end{isabellec}
\end{lemma}

Finally, we conclude the satisfaction of \isacodeinline{strong\_noninfluence} on our specification and all other information flow security properties according to the inference framework.

\section{Results and Discussion}
\label{sec:reslt_disc}

\subsubsection{Evaluation}
We use Isabelle/HOL as the specification and verification system for separation kernels. 
The proofs of information flow security in our specification are conducted in the structured proof language \emph{Isar} in Isabelle, allowing for proof text naturally understandable for both humans and computers. All derivations of our proofs have passed through the Isabelle proof kernel.

\begin{table}[t]
\centering
\scriptsize
\caption{Specification and Proofs Statistics} 
\begin{tabular} {|c|c|c|c||c|c|c|c|}
\hline
\multicolumn{4}{|c||}{\textbf{Specification}} & \multicolumn{4}{c|}{\textbf{Proofs}} 
\\ \hline
Item & \tabincell{c}{\# of function \\ /definition} & LOC & PM & Item & \tabincell{c}{\# of lemma \\ /theorem} & LOP & PM
\\\hline
Execution model & 32 & $\sim$ 200 & \multirow{3}{*}{2} & Inference Framework & 61 & $\sim$ 1000 & \multirow{3}{*}{6}
\\ \cline{1-3} \cline{5-7}
\multirow{2}{*}{Event Specification} & \multirow{2}{*}{68} & \multirow{2}{*}{$\sim$ 800} & & Correctness & 33 & \multirow{2}{*}{$\sim$ 6000} & 
\\ \cline{5-6}
& & & & Security & 123 & &
\\\hline
\textbf{Total} & 100 & $\sim$ 1000 & 2 & \textbf{Total} & 217 & $\sim$ 7000 & 6
\\\hline
\end{tabular}
\label{tbl:stat}
\end{table}

The statistics for the effort and size of the specification and proofs are shown in {\tableprefix} \ref{tbl:stat}.
We use 100 functions/definitions and $\sim$ 1000 lines of code (LOC) of Isabelle/HOL to specify the execution model and event specification. 217 lemmas/theorems in Isabelle/HOL are proved using $\sim$ 7000 lines of proof (LOP) of Isar to ensure the information flow security of our specification. The work is carried out by a total effort of roughly 8 person-months (PM).

\subsubsection{Validating and Fixing Covert Channels in ARINC 653}
When proving the satisfaction of unwinding conditions on the events, we find some security flaws, i.e., covert channels to leak information, in ARINC 653. 

\emph{Covert Channel 1: queuing mode channel-based communication}.
If there is a queuing mode channel from partition \isacodeinline{a} to \isacodeinline{b} and no other channels exist, then it is secure that \isacodeinline{a $\leadsto$ transmitter}, \isacodeinline{transmitter $\leadsto$ b}, \isacodeinline{transmitter {\isasymsetminus}{\isasymleadsto} a} and \isacodeinline{b {\isasymsetminus}{\isasymleadsto} transmitter}. 
In fact, these security policies are violated in ARINC 653.
Firstly, when \isacodeinline{a} sends a message by invoking \isacodeinline{Send\_Queuing\_Message} service of ARINC 653, the service returns \isacodeinline{NOT\_AVAILABLE} or \isacodeinline{TIMED\_OUT} when the buffer is full, and returns \isacodeinline{NO\_ERROR} when the buffer is not full. However, the full/empty status of the buffer in the port can be changed by message transmission executed by the transmitter. Thus, the \isacodeinline{local\_respect} property is not preserved on \isacodeinline{Send\_Queuing\_Message} service, and \isacodeinline{transmitter {\isasymsetminus}{\isasymleadsto} a} is violated. Secondly, due to no message loss required by ARINC 653, the transmitter cannot transmit a message on a channel when the destination queuing port is full. However, the full status of the destination port can be changed by \isacodeinline{Receive\_Queuing\_Message} service executed by partition \isacodeinline{b}. Thus, the \isacodeinline{local\_respect} property is not preserved on the event of message transmission, and \isacodeinline{b {\isasymsetminus}{\isasymleadsto} transmitter} is violated. To avoid this covert channel, we allow message loss when sending messages to a queuing port or transmitting message on a queuing mode channel.

\emph{Covert Channel 2: Create\_Sampling\_Port and Create\_Queuing\_Port services}. This is a potential covert channel. It is dependent on the concrete implementation of ARINC 653 and can be avoided by careful designs. 
In ARINC 653, the service \isacodeinline{Create\_Sampling\_Port} and \isacodeinline{Create\_Queuing\_Port} create a port and return a new unique identifier assigned by the kernel to the new port. In the initial specification, we use a natural number to maintain this new identifier. This number is initially assigned to one and increased by one after each port creation. We find in this design that the number becomes a covert channel that can flow information from any partition to another, and the two events do not preserve the \isacodeinline{weak\_step\_consistent} property. This covert channel can be avoided by assigning the port identifier to each port during system initialization or in the system configuration. 

\subsubsection{Validating and Fixing Covert Channels in Open-source Implementations}
We have manually validated the found covert channels in two open-source separation kernels, i.e., XtratuM and POK. Covert channels are found when we validate these two implementations.

The version of XtratuM we validate is v3.7.3 for SPARC v8 architecture. Unlike that there is one buffer for each queuing port in ARINC 653, XtratuM uses one shared buffer between the source port and the destination port of a queuing mode channel as a transmitter. 
If the buffer is not full, the hypercall \emph{SendQueuingPort} inserts the message into the buffer and notifies the receiver; whilst if the buffer is full, \emph{SendQueuingPort} immediately returns \emph{XM\_OP\_NOT\_ALLOWED}. The hypercall \emph{ReceiveQueuingPort} has a similar design. Thus, the found covert channel 1 exists in XtratuM.
The way to avoid this security flaw is to redesign the hypercall \emph{SendQueuingPort} to lose the message and return \emph{XM\_OK} when the buffer is full.

The version of POK we validate is the latest one released in 2014. Different from XtratuM, POK has a transmitter to transfer messages from a source port to a destination port of a channel. POK blocks processes to wait for resources. If the buffer is not full, the syscall \emph{pok\_port\_queueing\_send} inserts the message into the buffer; whilst if the buffer is full and $timeout = 0$, it immediately returns \emph{POK\_ERRNO\_FULL}. \emph{pok\_port\_transfer} responds for transmitting messages from a source port to a destination one and returns \emph{POK\_ERRNO\_SIZE} when the destination port has no available space to store messages. Thus, the found covert channel 1 exists in POK.
The way to avoid this security flaw is to allow message loss or block the calling process until the port buffer is not full in the syscall \emph{pok\_port\_queueing\_send}. 

When creating a port, XtratuM and POK use the index of the port in the port array as the new identifier. Thus, they do not have the covert channel 2.

\subsubsection{Discussion} The reusability of formal specification and proofs can largely alleviate the enormous efforts needed when others enforce information flow security on separation kernels. Our formal specification can be refined to the concrete specification of separation kernels.
In the concrete specification, new variables and events may be introduced and some events in this paper may be refined. 
The state equivalence in our specification is sufficient for the abstract and concrete specification of the channel-based communication. Therefore, the new variables in the concrete specification do not change the definition of state equivalence, and thus the new variables and new events manipulating these variables do not break the information flow security of the concrete specification. Information flow security properties in this paper can be preserved on refinement of events of the channel-based communication according to the conclusion in \cite{Murray12}.
Due to the reusability of the formal specification, the inference framework and the security proofs in this work are also reusable for the concrete specification. 

\section{Related Work and Conclusions}
\label{sec:rel_work}

\subsubsection{Information Flow Security}
Information flow security \cite{sabelfeld03} has attracted many research efforts in recent years. 
State-event based noninterference \cite{rushby92} is usually chosen for verifying general purpose operating systems and separation kernels \cite{Murray12}. Language-based information flow security was generalized to arbitrary multi-domain policies in \cite{von04} as a new state-event based notion nonleakage. Oheimb \cite{von04} also combined the classical noninterference and nonleakage as the notion noninfluence. These properties have been instantiated for operating systems in \cite{Murray12} and formally verified on seL4 \cite{Murray13}.
In our work, all of these properties and their variants are defined in our specification. We also propose an inference framework to clarify the implications between these properties.

\subsubsection{Formal Specification and Verification of Separation Kernels}
Formal methods have been widely applied on separation kernels in recent years \cite{craig07,velykis10,Verb14,Klaus15,Heitmeyer08,Richards10,Wilding10,freit11,Murray13,Dam13,sanan14}.
An overview is available in \cite{zhao15a}.
An Isabelle/HOL specification for a generic separation kernel was published by EURO-MILS project \cite{Verb14}. They provided an abstraction specification for Controlled Interruptible Separation Kernels (CISK), instantiated it to a separation kernel model, and then applied them on the PikeOS separation kernel \cite{Klaus15}. The Isabelle/HOL specification of seL4 was extended to a separation kernel specification in \cite{Murray13}. 
Formal specification in our work provides a detailed model for ARINC 653 channel-based communication, which is not covered in related work. In particular, there is no concrete communication actions in specification of \cite{Verb14}. The IPC syscalls in seL4 \cite{Murray13} and PikeOS \cite{Klaus15} are very different from ARINC 653 channel-based communication.

\subsubsection{Formalization and Verification of ARINC 653}
Formalization and verification of ARINC 653 have been considered in recent years, such as formal specification of ARINC 653 architecture \cite{Oliveira12}, modeling ARINC 653 for model driven development of IMA applications \cite{Delan10}, and verification of application software on top of ARINC 653 \cite{de11}. In \cite{zhao15}, the system functionalities and all service requirements in ARINC 653 have been formalized in Event-B, and some inconsistencies have been found in the standard. These works aim at safety of separation kernels or applications. Our work is the first to conduct a formal security analysis of the ARINC 653 standard.

\subsubsection{Conclusions and Future Work}
\label{sec:conclude}
The long-term goal of our project is to verify security of separation kernels on source code level.
In this paper, we presented a case study of applying Isabelle/HOL to formally specify and verify separation kernels with ARINC 653 channel-based communication. We provided a formal specification with mechanically checked proofs that is totally free of covert channels and therefore provided information flow security for high assurance systems. We revealed covert channels in ARINC 653 and validated their existence in XtratuM and POK. 
Our specification is reusable for subsequent specification refinement and development of implementations. The proofs in this work can alleviate the verification efforts on information flow security. 
In the next step, we will develop a formal specification of separation kernels supporting multi-core and the specification in this paper will be revised. Due to the kernel concurrency between cores, we will find a feasible way to verify multi-core separation kernels. 


\bibliographystyle{splncs03}
\bibliography{ref}

\end{document}